\documentclass[graybox]{svmult}

\usepackage{type1cm}        
\usepackage{makeidx}         
\usepackage{graphicx}        
\usepackage{multicol}        
\usepackage[bottom]{footmisc}

\usepackage{newtxtext}       %
\usepackage[varvw]{newtxmath}  
\usepackage{graphicx}
\usepackage{bookmark}
\usepackage{times}
\usepackage{xcolor}
\usepackage{verbatim}
\usepackage[utf8]{inputenc}
\usepackage{hyperref}
\usepackage{listings}
\usepackage{nameref}
\usepackage{svg}
\usepackage{calc}
\usepackage{newfloat}
\usepackage{makecell}
\usepackage{minted}
\usepackage{subfig}
\usepackage{multirow}
\usepackage{float}
\usepackage{bbding}
\restylefloat{table}

\begin{document}

\title*{Implementing Agents in JavaScript}
\author{Timotheus Kampik\orcidID{0000-0002-6458-2252}}
\institute{Department of Computing Science \at Umeå University,  901 87 Umeå, Sweden, \email{tkampik@cs.umu.se}}
%
%
\maketitle
\abstract{
This chapter gives an introduction to agent-oriented programming in JavaScript.
It provides an example-based walk-through of how to implement abstractions for reasoning loop agents in \emph{vanilla} JavaScript.
The initial example is used as a stepping stone for explaining how to implement slightly more advanced agents and multi-agent systems using \emph{JS-son}, a JavaScript library for agent-oriented programming.
In this context, the chapter also explains how to integrate reasoning loop agents with generative AI technologies---specifically, large language models.
Finally, application scenarios in several technology ecosystems and future research directions are sketched.}
\sloppy
%
%
\section{Introduction}
\label{sec:intro}
%
Over the past decades, JavaScript has evolved from a language for simple animations in Web pages to one that is powering not only modern, complex front-ends of many applications in Web browsers, on desktop machines and mobile applications, but also substantial back-end systems, utilizing run-time environments such as Node.js.
The language's importance and, indeed, its outright dominance in some of the most prevalent technology ecosystems, makes using JavaScript---or languages compiling to it---a necessity in many software engineering contexts.
Despite its popularity, which is also evidenced by programming language rankings~\cite{programminglanguages}, JavaScript still receives relatively little attention in contexts that are heavily academically influenced, possibly due to its perceived ugliness and its weak type system.
This may explain why JavaScript is somewhat under-explored as a technology for implementing agents and Multi-Agent Systems (MAS).
Still, as we demonstrate in this chapter, it is not only relatively easy to implement reasoning loop agents and MAS using JavaScript, its flexible and functional nature also caters to using the notion of an agent as a fundamental abstraction.
We give a brief practical overview of the Agent-Oriented Programming (AOP) paradigm in Section~\ref{sec:paradigms}, followed by a conceptual overview of how AOP can be utilized for implementing agents in JavaScript (Section~\ref{sec:overview}).
To emphasize simplicity, we first implement agents from scratch, entirely in vanilla JavaScript, focusing on belief-plan deliberation, before we specify more elaborate agents using the JS-son library~\cite{DBLP:conf/emas/KampikN19} (Section~\ref{sec:agents}).
We then move from single agents to MAS and agent-based simulations (Section~\ref{sec:mas}).
Highlighting broad application potential, we discuss several use cases with different deployment targets (Section~\ref{sec:applications}).
Finally, we conclude by sketching usage scenarios and future research directions, and by appealing to students, researchers, and practitioners alike to advance both the foundations and application of JavaScript agents (Section~\ref{sec:conclusion}).
Links to code examples and other resources are provided in the appendix.
Note that while we relate to Large Language Models (LLMs) as integration components for reasoning loop agents, the concepts and design patterns we introduce are LLM-agnostic.

%
\section{Agent-Oriented Programming}
\label{sec:paradigms}
%
Agents are a (if not \emph{the}) fundamental abstraction of Artificial Intelligence (AI), which is sometimes described as the ``study of agents that receive percepts from the environment and perform actions''~\cite[p.viii]{russel2010}.
The importance of the notion of an \emph{agent} in both academic and practical computer programming contexts motivated researchers to coin the term \emph{Agent-Oriented Programming} (AOP), initially as a specialization of Object-Oriented Programming (OOP)~\cite{SHOHAM199351}.
In contrast to objects in OOP, agents in AOP are proactive (as well as reactive), and run a \emph{reasoning loop} process that revises their internal state---so-called \emph{beliefs}---based on percepts obtained from their environment to ultimately execute plans and decide on actions that are then relayed to the environment.
Several conceptualizations of reasoning loop agents have been proposed over the years; particularly prominent are so-called \emph{belief-desire-intention} agents that move from beliefs to actions by first considering what they would like to achieve as potentially mutually inconsistent \emph{desires}, before determining the \emph{intentions} that a given agent actually strives towards realizing~\cite{rao91a}.
Over the past three decades, a heterogeneous ecosystem of AOP tools and languages has emerged\footnote{For a recent survey of tools and platforms for programming agents and multi-agent systems, see~\cite{DBLP:conf/emas/BriolaFM23}.}.
Many languages designed or used for AOP are \emph{not} object-oriented: perhaps most notably, the \emph{AgentSpeak} language and its dialects are Prolog-like, i.e., logic programming is used as a paradigm orthogonally to AOP~\cite{DBLP:conf/maamaw/Rao96,Bordini:2007:PMS:1197104}.
In order to facilitate applicability, researchers have integrated AgentSpeak into mainstream programming language ecosystems, e.g., for Java~\cite{Bordini:2007:PMS:1197104,BOISSIER2013747} and Kotlin~\cite{DBLP:conf/eumas/BaiardiBCP23}, and orthogonally into distributed environments such as micro service-based systems~\cite{DBLP:conf/emas/ONeillC23} and the Web~\cite{DBLP:conf/emas/CiorteaBR18}.
However, it is still a challenge for practitioners to adopt AOP~\cite{engineering-gsi-article-2019}, arguably partially because the combination of AOP and logic programming---two paradigms that are not intuitively understood by typical software engineers---leads to an unnecessarily steep learning curve.
Prominently, in the wake of the hype around Large Language Models (LLMs), proposals for designing so-called \emph{AI agents} that integrate LLMs with reasoning loops in the broader sense (cf.~\cite{llmagentsurvey} for a survey), largely ignore AOP and its rich history.

The aforementioned challenges motivate the approach to AOP that we take in this chapter: the implementation of AOP design patterns as re-usable abstractions in and for a mainstream programming language that is of crucial importance for a broad range of application scenarios.
We argue that the focus on JavaScript is particularly interesting, not only because of its tremendous real-world prevalence, but also because it is exactly \emph{not} one of the eloquent and theoretically well-understood languages that academics prefer to study.
In other words, we turn the approach that typical AOP approaches take on its head, starting with a mainstream programming language and adopting AOP-like abstractions to the extent we find practically useful.
While this means that we sacrifice some of the formal and philosophical elegance of traditional AOP, it allows us to position AOP in a way that may be more intuitive for practitioners, as well as for students and researchers that primarily work with mainstream programming languages.
To this aim, we also present our AOP approach to JavaScript both generically and as part of a somewhat \emph{lean} library, with little code overhead and no external dependencies: while the library makes it easier to get started with implementing agents in JavaScript, serious applications may require custom implementations of agent-oriented abstractions; for the latter scenario, this chapter provides a conceptual overview as well as technical descriptions relating AOP directly to JavaScript.

\section{Conceptual Overview}
\label{sec:overview}
\begin{figure}[ht!]
\centering
\includegraphics[width=8cm]{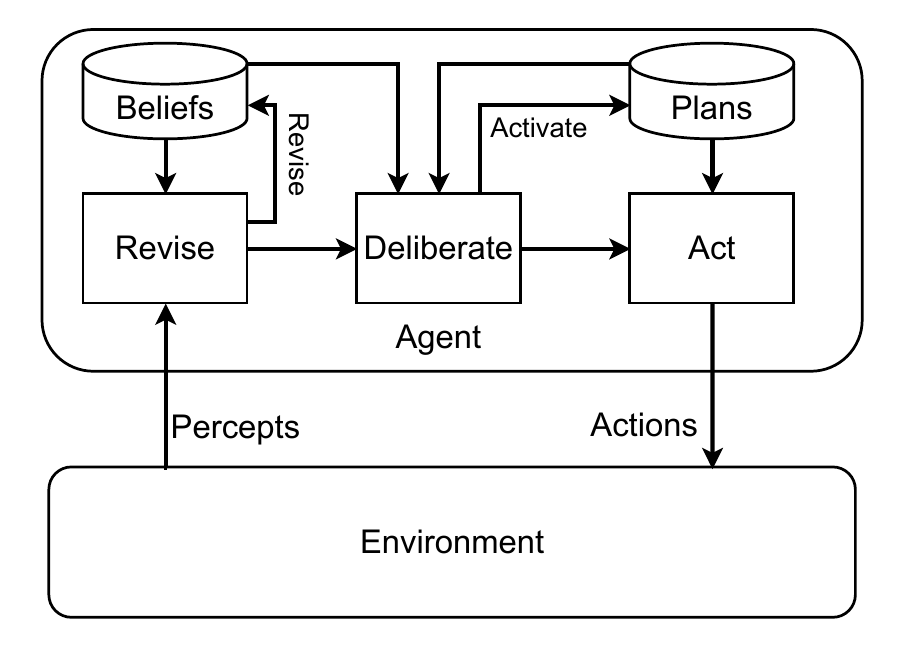}
\caption{A reasoning loop agent, interacting with a black-box environment (unlabeled arrows represent information transfer).}
\label{fig:reasoning-loop}
\end{figure}

%
Before moving to hands-on implementation tutorials, we provide an overview of relevant abstractions for designing reasoning loop agents, already hinting at how these can potentially be implemented in JavaScript, and giving basic application intuitions.
As a starting point, let us consider the reasoning loop architecture sketched in Figure~\ref{fig:reasoning-loop}.
As we can see in the figure, our reasoning loop agent perceives its environment in that it receives \emph{percepts}. The percepts are then, in conjunction with its existing \emph{beliefs}, applied to \emph{revise} the belief base.
Based on the revised beliefs, the agent \emph{deliberates} to determine which of its \emph{plans} are activated (based on a given plan's \emph{head}, see below) to then execute the \emph{bodies} of activated plans, yielding \emph{actions}.
Subsequently, the environment processes the actions, returning another set of percepts, and the cycle continues.

The selection of the abstraction is, in parts, opinionated and based on an engineering intuition of what is and is not useful for practical applications.
The two overarching notions of \emph{agents} and \emph{environments} are obvious and uncontroversial choices.
\begin{description}
    \item[\textbf{Agents.}] As described above, agents process percepts, maintain beliefs and plans, activate plans based on current beliefs, and determine actions that are to be registered with the environment (by executing plans). \\
    \emph{Example.} An agent tasked with handling a person's travel expenses and claims.
    \item[\textbf{Environments.}] The environment processes actions of agents and generates agents' percepts, thus also orchestrating interactions between agents. \\
    \emph{Example.} A middle-ware integrating our travel expense agent with different systems.
\end{description}
\newpage
Both \emph{agents} and \emph{environments} are proper JavaScript objects, i.e., neither merely simple JSON objects nor functions:
because agents and environments carry state, they cannot (at least not conveniently) be functions;
because they must actively process inputs and generate outputs, they cannot be plain JSON objects (in the sense of the Internet Engineering Taskforce spec~\cite{rfc8259}), which only allow for \emph{representation} but not for \emph{reasoning}.

As the chapter focuses on the implementation of \emph{agents} in JavaScript, we are mainly interested in \emph{agent internal} abstractions and relegate---colloquially speaking---the role of the environment to plumbing and piping.
Accordingly, we focus on agent internals in this initial conceptual overview\footnote{Still, some important environment characteristics are discussed in Section~\ref{sec:mas}.}.
Let us first introduce the key notions of agent-internal state (as well as the state update that an agent receives from the enviornment), primarily dividing between \emph{percepts and beliefs}, i.e., the agent's world model, and \emph{plans} and related notions that are the foundation of the agent's specific reasoning behavior.
\begin{description}
    \item[\textbf{Percepts and Beliefs.}] Percepts and beliefs determine an agent's model of the world: the agent receives percepts from the environment and updates its beliefs considering these percepts, to an extent the agent decides autonomously.
    In the JS-son library, percepts are, intuitively, \emph{incoming} beliefs that agents merge with the already existing internal beliefs when applying the belief revision functions: i.e., percepts are not modeled using a separate abstraction, but instead are belief objects that an agent perceives and then revises before adding them to its belief base.
    In its most basic form, percepts and beliefs are static and can be modeled as (key, value)-tuples in a JSON object. However, in advanced scenarios beliefs may, in addition, have numeric priority values helping decide which of two conflicting beliefs should be adopted, or are functions themselves that dynamically infer values given other beliefs. \\
    \emph{Example.} Expense claims that need to be handled and their meta-data, as well as documents that serve as proofs of actual spending.
    \item[\textbf{Plans (with Head and Body), Goals, and Actions.}] Plans are \emph{(activation condition, action inference function)}-tuples, allowing the agent to dynamically and autonomously decide what actions to take. Here, the activation condition is the \emph{head} of the plan in the form of a Boolean function that takes as input an agent's beliefs. One may say that an activated plan models a \emph{goal}. The body of a plan determines the plan's action(s), again potentially depending on the beliefs of the agents. \\
    \emph{Example.} Whether to gather more information about a claim and whether to submit one (heads), as well as where and how to gather information/submit claims (bodies).
\end{description}
Given its internal state and perception of the environment, an agent's reasoning loop consists of the following key steps.
\begin{description}
    \item[\textbf{Belief Revision} (\emph{revise})\textbf{.}] The agent's belief revision function takes a set of incoming beliefs as perceived by the agent (hence: \emph{percepts}), as well as the beliefs the agent currently holds, and returns a set of beliefs that forms the agent's new belief base. The key idea is that the agent has \emph{autonomy} over its beliefs, using the belief revision function to decide how to process what it perceives in its environment. In Figure~\ref{fig:reasoning-loop}, belief revision is the \emph{revise} step of the reasoning loop. \\
    \emph{Example.} Augmenting and revising the meta-data of a previously rejected claim. 
    \item[\textbf{Plan Deliberation} (\emph{deliberate} and \emph{act})\textbf{.}] The agent's plan deliberation function first checks which plans in an agent's plan library are active and then executes active plans, yielding \emph{actions}. In Figure~\ref{fig:reasoning-loop}, plan deliberation is covered by the \emph{deliberate} and \emph{act} steps: \emph{deliberate} activates plans based on their heads; \emph{act} determines actions that are then registered with the environment, based on the bodies of activated plans. \\
    \emph{Example.} Deciding what to request and submit, and where. 
\end{description}

Table~\ref{agent-abstractions} provides an overview of the AOP abstractions and how they can be modeled in JavaScript.
Also, the table indicates whether the corresponding abstraction is an explicit, first-class abstraction in our AOP approach (and thus in the JS-son library) or rather modeled implicitly, as part of another abstraction.
Let us highlight that the table does not reflect the details of the function signatures and object definitions of the JS-son library, as it is aimed at a higher level of conceptual abstraction. Technical details are available in the JS-son documentation\footnote{Cf. Appendix Item~\ref{app:js-son-doc}.}. Notably, we skip the additional capabilities for goal and desire deliberation that JS-son supports, in order to increase ease-of-understanding and, ultimately, applicability.
\begin{table}
     \def\arraystretch{1.5}
     \setlength\tabcolsep{10pt}
     \caption{First-Class (explicit) and second-class (implicit) abstractions of JS-son JavaScript agents; $f$ stands for \emph{function}.}
     \label{agent-abstractions}
     \centering
     \begin{tabular}{ l | l | l | l }
          \hline
          \textbf{Abstraction} & \textbf{Modeled As}  & \textbf{First-Class?}  & \textbf{Covered By} \\ \hline
          Agent & JavaScript Object & \Checkmark &  \\
          Environment & JavaScript Object & \Checkmark &  \\
          Belief & JSON (Key, Value)-tuple & \Checkmark &  \\
          Percept & JSON (Key, Value)-tuple & \XSolidBrush & Belief \\ 
          Plan & $f$-tuple: (Head, Body)  & \Checkmark & \\ 
          Plan Head & $f$: Beliefs $\rightarrow$ Boolean   & \Checkmark & \\ 
          Plan Body & $f$: Beliefs $\rightarrow$ Actions  & \Checkmark & \\ 
          Goal & $f$: Beliefs $\rightarrow$ Boolean & \XSolidBrush & Plan Head \\
          Action & JSON Object & \XSolidBrush & Plan \\
          Belief Revision & $f$: Beliefs, Percepts $\rightarrow$ Beliefs & \Checkmark &  \\
          Plan Deliberation & $f$: Beliefs, Plan $\rightarrow$ Boolean & \XSolidBrush & Agent  \\
          \hline
     \end{tabular}
\end{table}
%

\section{Implementing Agents}
\label{sec:agents}
%
JavaScript is a language that adopts both functional and object-oriented notions as key abstractions.
When implementing JavaScript agents, we can utilize this dichotomy: roughly speaking, an agent's reasoning loop can be defined as a composition of (somewhat impure) functions, whereas the beliefs it maintains are objects.
From the belief base, the agent can draw inferences and turn them into actions, again utilizing functions that are maintained in a plan library object.
This intuition yields a \emph{belief-plan} agent, i.e., a rather simple reasoning loop agent.
It is relatively straight-forward to implement a belief-plan agent and its environment from scratch, which we do in Subsection~\ref{subsec:from-scratch}. In Subsection~\ref{subsec:llm-agent}, we make use of the JS-son library to implement a slightly more involved agent that utilizes an LLM.
The purpose of starting with an example that implements an agent from scratch is to relay a nuanced understanding of agent programming in JavaScript, avoiding any potential obfuscation by non-standard abstractions.
The tutorial can help programmers working with ``ordinary'' JavaScript, alongside mainstream libraries and frameworks, to think in an agent-oriented manner and to ultimately implement their own agent-oriented abstractions.
However, readers who wish for a more colloquial tutorial that helps them get started with agent-programming with re-usable abstractions in an easy-going manner may jump to Subsection~\ref{subsec:llm-agent} right away.
\subsection{A Simple Agent in Vanilla JavaScript}
\label{subsec:from-scratch}
As a minimal example, consider a \emph{porter} agent\footnote{Cf. Appendix Item~\ref{app:jason-room}.} that observes whether a door is locked or not, and opens and closes the door based on incoming requests, given the requested state of the door is currently not satisfied.

First, we model the \emph{beliefs} of the agent.
\begin{minted}[fontsize=\scriptsize]{js}
    const beliefs =
        { door: { locked: true },
          requests: [] }
\end{minted}
As we can see, the initial state of the belief base models that the door is currently locked, and that no requests are to be handled.
Let us now specify the agent's plans: \emph{i)} if the door is not locked and a request to lock the door exists (\emph{head}) then the agent should act to lock the door (\emph{body}); \emph{ii)} if the door is locked and a request to unlock the door exists (\emph{head}) then the agent should act to unlock the door (\emph{body}).
\begin{minted}[fontsize=\scriptsize]{js}
    const plans = [
        { head:
            beliefs =>
                !beliefs.door.locked &&
                beliefs.requests.includes('lock'),
         body: () => 'lock' },
        { head: beliefs =>
            beliefs.door.locked &&
            beliefs.requests.includes('unlock'),
         body: () => 'unlock' }
    ]
\end{minted}
With beliefs and plans specified, we can move on to implement the agent's reasoning loop.
We start by implementing the belief revision function, which updates the agent's belief about the state of the door, removes requests that have been successfully executed, and adds new requests that have come in.
\begin{minted}[fontsize=\scriptsize]{js}
    const revise = (beliefs, percepts) => {
        return {
            door: percepts.door,
            requests: beliefs.requests.filter(
                request =>
                    !percepts.executions.includes(request)
            ).concat(percepts.requests)
        }
    }
\end{minted}
Next, we implement the plan deliberation function.
The function first determines which bodies of the plans should be activated (\emph{deliberate}) to then execute these bodies (\emph{act}).
\begin{minted}[fontsize=\scriptsize]{js}
    const deliberate = (beliefs, plans) => {
        return activatedBodies = plans.filter(
            plan =>
                plan.head(beliefs)
        ).map(plan => plan.body(beliefs))
    }
\end{minted}
Finally, we can complete the agent, by \emph{i)} assigning beliefs and plans, as well as the revision and deliberation functions to an agent object, and \emph{ii)} implementing a \verb|run| function that triggers first belief revision and then plan deliberation.
\begin{minted}[fontsize=\scriptsize]{js}
    const agent = {
        beliefs,
        plans,
        revise,
        deliberate,
        get run() {
            this.beliefs = this.revise(this.beliefs, percepts)
            return this.deliberate(
                this.beliefs,
                this.plans
            )
        }
    }
\end{minted}
We can observe that we need to make use of a \verb|get| function to access object-level properties and that \verb|percepts| will need to be specified outside of the scope of the agent object or any of its functions.
This indicates that providing generic abstractions in the form of a library may be useful.

We still need to implement the environment that provides our agent with percepts and processes its actions.
%
\begin{minted}[fontsize=\scriptsize]{js}
    const environment = {
        agent,
        state: {
            door: { locked: true },
            requests: []
        },
        get run() {
            [...Array(steps).keys()].forEach(
                step => {
                    this.state.requests =
                        Math.random() < 0.5 ?
                            ['lock'] :
                            ['unlock']
                    percepts = this.state
                    const exec = this.agent.run
                    const actions = exec ? exec : []
                    if (actions.includes('lock'))
                        this.state.door.locked = true
                    if (actions.includes('unlock'))
                        this.state.door.locked = false
                    this.state.executions = actions
                }
            )
        } 
    }
\end{minted}
Our environment processes all agent actions by changing the state of the door accordingly and adds pseudo-random requests to either lock or unlock the door, independently of the door's current status.
In this way, the environment primitively simulates the presence of additional agents.
Note that in this simple example, our environment is a \emph{single agent environment}, i.e., it is not programmed to handle more than one agent.
We again exploit the \verb|get| syntax in a way that is arguably not idiomatic.

We can now execute the environment, first specifying the number of environment-agent interactions (\verb|steps|) it should run.
\begin{minted}[fontsize=\scriptsize]{js}
    const steps = 20
    environment.run
\end{minted}
When running the example, we see that our porter agent processes the requests it receives from the environment as intended, generating the desired actions, which are then further processed accordingly by the environment\footnote{Cf. Appendix Item~\ref{app:vanilla-example}.}.
When debugging nuances of the example, we observe that nonsensical requests (the locking of locked doors and the unlocking of unlocked doors) ``pile up'' until they can be meaningfully addressed.

As we have seen, it is relatively straightforward to implement a simple reasoning loop agent in JavaScript.
Still, even in this simple example, in which a purely reactive agent handles a straightforward task whose completion merely requires a single reasoning cycle, we see that we can implement abstractions that \emph{i)} guide developers so that they implement agents in a somewhat idiomatic manner and \emph{ii)} reduce overhead and duplicate work during the implementation.
Below, we make use of such abstractions to implement a slightly more involved agent.

\subsection{An LLM Agent Based on Reusable Abstractions}
\label{subsec:llm-agent}
We implement a reasoning loop agent utilizing an LLM in the following scenario:
Our agent acts on behalf of a student who has forgotten to do their homework and aims to submit an excuse request to the environment.
The environment either accepts or rejects the request and gives feedback regarding why a request was denied.
As the student has several attempts to submit the request, their agent adjusts the excuse request for subsequent attempts based on feedback that the environment has provided.

We first implement the agent's beliefs, by instantiating JS-son \verb|Belief| objects.
\begin{minted}[fontsize=\scriptsize]{js}
    const beliefs = {
        ...Belief('rejectExps', []),
        ...Belief('excuseAccepted', false),
        ...Belief('name', 'Bart'),
        ...Belief('teacherName', 'Edna Krabappel')
    }
\end{minted}
In our case, the result is roughly the same as when specifying beliefs as JavaScript objects.
However, in advanced scenarios, JS-son beliefs have in-built support for specifying revision behavior, such as priorities and inter-belief dependencies, where the value of one belief is inferred from the value(s) of one or several other beliefs.

For the integration with the LLM, we assume a \verb|model| object with a \verb|generatePrompt| function that takes a text as input and returns a \emph{promise} (for asynchronous function call handling), ultimately yielding text output.
In our implementation, we rely on the \emph{Gemini} LLM~\cite{team2023gemini}, developed and provided by Google.

We specify a prompt template for instantiating a prompt based on the agent's beliefs, considering the agent's name (the name of the person the agent represents), the name of the teacher, and information about why previous excuse attempts (if any) have failed.
\begin{minted}[fontsize=\scriptsize]{js}
const genPrompt = beliefs =>
    `Can you write a charming yet convincing excuse
    for a student who forgot their homework?
    The names of teacher and student are
    ${beliefs.teacherName}, and ${beliefs.name},
    respectively (i.e., sign the excuse with ${beliefs.name}).
    
    Consider the following feedback
    received from past rejected excuses:

    ${beliefs.rejectExps.map(exp => `• ${exp}`).join('\n')}`
\end{minted}
The prompt template is called in the body of the agent's single plan.
The body is activated (by the plan's head) only if no excuse has been accepted so far.
The body then constructs the excuse prompt and sends it to the LLM.
When the agent has received the generated excuse, it returns the excuse to the environment.
\begin{minted}[fontsize=\scriptsize]{js}
Plan(
    beliefs => !beliefs.excuseAccepted,
    async beliefs => {
        const prompt = genPrompt(beliefs)
        const excusePromise =
            await model.generateContent([prompt])
        const excuse = excusePromise.response.text()
        return excuse
    }
)
\end{minted}
Above, we have instantiated a JS-son \verb|Plan| object; later, the agent will apply a default deliberation function, thus saving us some implementation overhead.

Next, we implement the agent's belief revision function.
JS-son's default revision function simply merges percepts into the agent's belief base, with percepts taking precedence over beliefs in case of conflicts.
However, for our agent we need a more specific belief revision function that \emph{i)} adds new rejection explanations (if novel) to our array of explanations and \emph{ii)} updates the \verb|excuseAccepted| belief based on the percepts.
\begin{minted}[fontsize=\scriptsize]{js}
    const reviseBeliefs = (beliefs, percepts) => {
        const rejectExps =
            percepts.rejectExp &&
            !beliefs.rejectExps.includes(percepts.rejectExp) ?
            Array(...beliefs.rejectExps, percepts.rejectExp) :
                beliefs.rejectExps
        return percepts.excuseAccepted ? {
            ...beliefs,
            excuseAccepted: true,
        } : {
            ...beliefs,
            rejectExps
        }
    }
\end{minted}
With our beliefs, plan, and belief revision function, we can then instantiate a JS-son \verb|Agent| object.
\begin{minted}[fontsize=\scriptsize]{js}
    const agent = new Agent({ id: 'student', beliefs, [plan], reviseBeliefs })
\end{minted}
Note that the agent's runner function (\verb|next| in JS-son terms) is built in, i.e., we do not need to implement it explicitly.

As a prerequisite to implementing the environment, we need to specify the \verb|updateState| function, of which we merely provide a sketch, abstracting from the exact checks executed when determining whether an excuse is acceptable or not.
\begin{minted}[fontsize=\scriptsize]{js}
    const updateState = async(actions, _, currentState) => {
        const excuse = await actions[0]
        if(!excuse)
            return { ...currentState, rejectExp: ''}
        const state = {
            excuseAccepted: false
        }
        if (...) { // apply rule to analyze excuse text
            state.rejectExp = ... // explain rule violation
        } else if (...) { // apply another rule to analyze excuse text
            state.rejectExp = ... // explain another rule violation
        } else {
            state.excuseAccepted = true
        }
        return state
    }
\end{minted}
As we can see, the function essentially applies a set of rules to assess whether the provided excuse is acceptable or not and, in case of a rule violation, provides an explanation for rejection (for the first rule that is violated, which terminates rule execution).
Then, the state of the environment is updated accordingly (which implicitly leads to a corresponding update of the agent's percepts). 

Finally, we specify the environment, using the agent, state update function, and an object defining the initial state (excuse not accepted and no explanations for rejections provided so far).
\begin{minted}[fontsize=\scriptsize]{js}
    const environment = new Environment(
        [agent],
        {   excuseAccepted: false,
            rejectExp: ''   },
        updateState
    )
    setInterval(() => environment.run(1), 3000)
\end{minted}
Again, we do not explicitly implement a runner function, as this is provided by default by JS-son's \verb|environment| object.
We run the environment in timed intervals.
This reduces load on the LLM service API, but also smoothens the execution of a JS-son reasoning loop with asynchronous behavior---in contrast to asynchronous environments, asynchronous behavior within reasoning loops is not supported in a straight-forward, stable manner by JS-son at the time of writing~\footnote{Cf. Appendix Item~\ref{app:llm-agent}.}.

While the scenario above has some simplistic aspects to it (such as the application of a few hard-coded rules for the analysis of natural language text), one can argue that it goes into the direction of somewhat realistic use cases.
For example, reasoning loop agents utilizing LLMs may be used in the future in the context of Robotic Process Automation (RPA), i.e., for the automation of simple form submissions and related tasks human users tend to find tedious, and LLMs may help brushing over some shortcomings in the corresponding form requirements and system interfaces.
At the same time, the example also hints at the fact that agents utilizing LLMs may be employed for socially questionable use-cases that do not lead to long-term sustainable progress.

%
\section{Implementing Multi-Agent Systems}
\label{sec:mas}
\begin{figure}[ht!]
\centering
\subfloat[Synchronous Interaction.]{
\label{fig:sync}
\includegraphics[width=0.5\columnwidth]{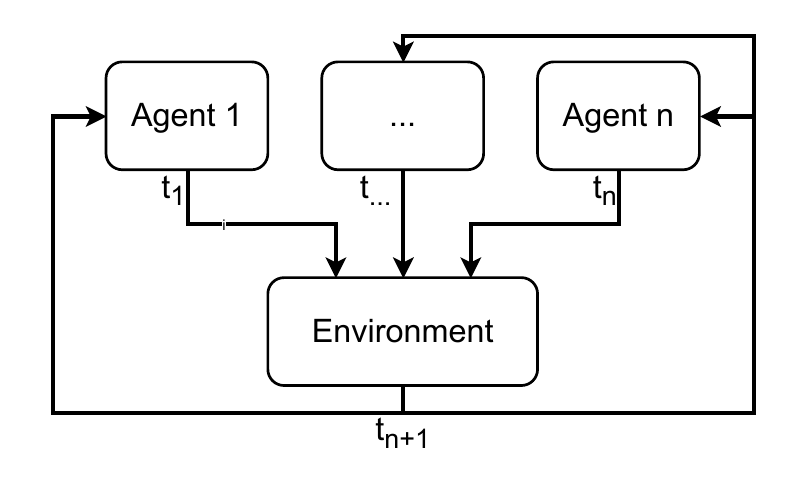}
}
\subfloat[Asynchronous Interaction.]{
\label{fig:async}
\includegraphics[width=0.5\columnwidth]{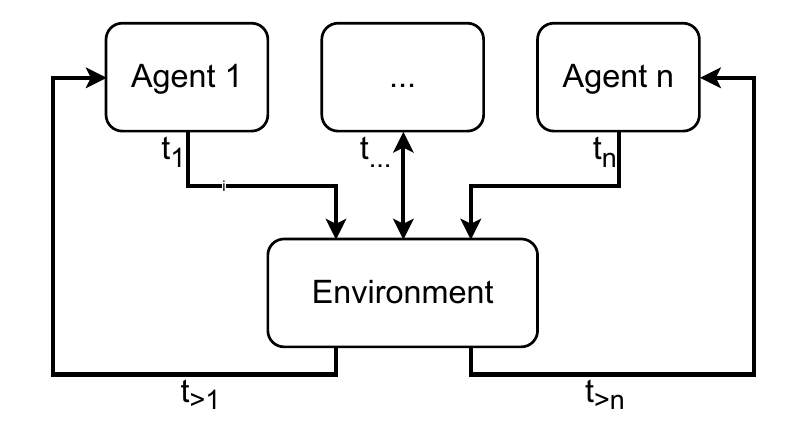}}
\caption{Multi-agent environments: synchronously scheduled environment, for round-robin based simulations or games vs. asynchronous environments for distributed MAS.}
\label{fig:envs}
\end{figure}

%
Obviously, agents typically do not only interact with their environment and passive artifacts therein, but also with other agents.
Hence, in multi-agent scenarios, the environment serves as a middle-ware, relaying communications between agents.
In some centralized scenarios, such as simple games and agent-based simulations, the environment may serve agents in a \emph{round robin} manner, first providing percepts to agents one-by-one, while always waiting for the termination of their reasoning loop to register their actions; the actions are then processed in a batch, i.e., the environment's state update is executed after all agents have acted (Figure~\ref{fig:sync}).
However, in more realistic applications, potentially (logically or physically) distributed agents must be able to act asynchronously. Then, the environment processes agent actions immediately when they are received, \emph{i.e.}, asynchronously across agents (Figure~\ref{fig:async}).
Below, we sketch two multi-agent examples, implementing centralized (synchronous, Subsection~\ref{subsec:centralized-mas}) and logically distributed (asynchronous, Subsection~\ref{subsec:dist-mas}) MAS, respectively.
For the sake of conciseness, we provide code examples to the extent they are illustrative and point to online material for the complete program code.

\subsection{Centralized Multi-Agent Systems}
\label{subsec:centralized-mas}
As our \emph{centralized} example, we implement a \emph{game of life} multi-agent system that runs in a Web browser.
Conway's \emph{Game of Life} (GoL) is a well-known example of a simplistic multi-agent simulation, formally grounded in cellular automata, illustrating how intriguing patterns can emerge from societies of agents that each individually follow very simple rules~\cite{gradner1970fantastic}.
Specifically, a GoL is a grid of agents, each of which is either \emph{active} or \emph{inactive} and changes, in discrete time, its status (from active to inactive or vice versa) based on its own current status and the statuses of its direct neighbors:
\begin{itemize}
    \item If the agent is already active, it will stay so if at least two and at most three of its neighbors are active.
    \item Otherwise, the agent turns active only if exactly three of its neighbors are active.
\end{itemize}
Intuitively, agents thrive in areas that are neither underpopulated nor overcrowded.

We can implement the agent behavior we have sketched above as a single plan.
\begin{minted}[fontsize=\scriptsize]{js}
    Plan(
            () => true,
            beliefs => {
              const neighborActivity =
                determineNeighborActivity(
                    beliefs.index, beliefs.activityArray
                )
              const isActive =
                beliefs.activityArray[beliefs.index]
              return (
                  isActive &&
                  neighborActivity >= 2 &&
                  neighborActivity < 4
              ) || neighborActivity === 3 ?
                { nextRound: 'active' } :
                { nextRound: 'inActive' }
    })
\end{minted}
Note that the plan is always active (its head necessarily evaluates to \verb|true|).
\newpage
The plan's body depends on the following beliefs:
\begin{itemize}
    \item The agent's index in the grid;
    \item The activity statuses of all agents (i.e., assigned by their indices in the grid).
\end{itemize}
Also, the agent makes use of the inference rule/function \\ \verb|determineNeighborActivity| that, given its index and the activity statuses of all agents, determines the number of active neighbors.

Given our plan, we can generate the entire grid of agents by using a simple function that assigns the agents' initial beliefs based on an \verb|initialActivity| array that specifies the initial state of the agents (for example, in order to guarantee the emergence of a specific pattern).
\begin{minted}[fontsize=\scriptsize]{js}
    const generateAgents = initialActivity =>
        initialActivity.map((value, index) => {
            const beliefs = {
                ...Belief('index', index),
                ...Belief('activityArray', initialActivity)
            }
        return new Agent(index, beliefs, {}, [plan])
    })
\end{minted}
Next, we need to specify the environment's behavior.
Here, it is crucial to assure that updates are propagated \emph{at the end of a given round} (cf. Figure~\ref{fig:sync}).
Accordingly, our environment distinguishes between \emph{previous} and \emph{next} agent activity statuses, right from the start when specifying the initial state.
\begin{minted}[fontsize=\scriptsize]{js}
    const generateState = initialActivity => ({
      previousActivity: initialActivity,
      nextActivity: []
    })
\end{minted}
Accordingly, the environment's state update is propagated only when the last agent in a grid has submitted its action.
\begin{minted}[fontsize=\scriptsize]{js}
    const updateState = (actions, agentId, currentState) => {
        const stateUpdate = { ...currentState }
        const agentActive =
            actions.some(action =>
                action.nextRound === 'active'
            )
        stateUpdate.nextActivity.push(agentActive)
        if (agentId == currentState.previousActivity.length - 1) {
            return {
                previousActivity: stateUpdate.nextActivity,
                nextActivity: []
            }
        }
        return stateUpdate
    }
\end{minted}
We also specify a state filter that exposes the agents only to the grid's ``previous'' status, i.e., not to the status that is currently undergoing an update.
\begin{minted}[fontsize=\scriptsize]{js}
    const stateFilter = state => (
        { activityArray: state.previousActivity }
    )
\end{minted}
Finally, we can generate the GoL's initial activity grid and instantiate the environment.
\begin{minted}[fontsize=\scriptsize]{js}
    const initialActivity = generateInitialActivity()
    const env = new Environment(
        generateAgents(initialActivity),
        generateState(initialActivity),
        updateState,
        render,
        stateFilter
    )
\end{minted}
With this we have finalized the specification of the GoL.
We can, for example, attach the MAS to a rendering function that produces/manipulates HTML elements and run it in a Web browser\footnote{Cf. Appendix Item~\ref{app:centralized-mas}.}.

\subsection{Distributed Multi-Agent Systems}
\label{subsec:dist-mas}
Next, we describe how to implement a simple logically distributed MAS, wherein agents interact asynchronously.
For this, we again go back to our initial \emph{porter} agent example.
This time, however, we implement an MAS and not a single-agent system, featuring two agents in addition to the porter:
\begin{itemize}
    \item A \emph{paranoid} agent, who always requests that the door be locked when it is unlocked;
    \item A \emph{claustrophobic} agent, who always requests that the door be unlocked when it is locked.
\end{itemize}
The porter agent always satisfies the latest request issued by either agent and thus iteratively unlocks and locks the door (until the MAS is terminated).
In our example, the porter and paranoid agents, as well as the environment, run on a Web socket server, whereas the claustrophobe agent runs on a client.
As some of the agents and the environment are executed in a single thread, we may technically consider the scenario a \emph{hybrid} case (a mix of Figures~\ref{fig:sync} and~\ref{fig:async}) instead of a fully distributed case.
In order to manage interactions with the claustrophobe, the environment maintains a \emph{shadow agent}, which serves as an interface to the client, abstracting away the details of the client and the agent running on it.
Like this, it is in principle possible to integrate JS-son MAS with agents implemented using other libraries and approaches.

As the agents are straightforwardly implemented, analogously to how the porter agent is implemented as described in Subsection~\ref{subsec:from-scratch} (though here using the JS-son library and not vanilla JavaScript), we focus below on what is required for managing the distributed nature of the MAS.
This means we assume that on server side we have:
\begin{itemize}
    \item The two local agents, \verb|porter| and \verb|claustrophobe|, who have the same simple set of beliefs (\verb|beliefs|) about the environment, cf. Subsection~\ref{subsec:from-scratch};
    \item The environment's state (\verb|state|) and state update function (\verb|update|);
    \item \verb|ws|, a Web socket connection object that allows us to listen to open connections (\verb|.on('open', ...)|) and messages (\verb|.on('message', ...)|), and send messages (\verb|.send(...)|).
\end{itemize}
What remain to be implemented are the shadow agent, environment-client communications, and the client agent.

Let us start with the former.
For this, we implement a higher-order function that generates our custom runner function that we will give to our shadow agent. The function the generator returns will send the belief update to the remote agent (instead of applying it to a locally running instance).
\begin{minted}[fontsize=\scriptsize]{js}
    const nextGen = agentId => function next (beliefs) {
        ws.send(JSON.stringify(beliefs))
        return global.actionRequests[agentId]
    }
\end{minted}
Now, we can instantiate the shadow agent, using JS-son's \verb|RemoteAgent| abstraction.
\begin{minted}[fontsize=\scriptsize]{js}
    const claustrophobe = new RemoteAgent(
        'claustrophobe',
        beliefs,
        nextGen(agentId)
    )
\end{minted}
Let us move to environment-client communications.
For this, we first implement a higher-order function generating a \verb|runner| for the environment. Because we need the Web socket client to be available in the context of this function, we create the 3rd-order function \verb|runnerGen|. The \verb|runner| higher-order function provides a context in which the environment's \verb|run| function is executed. In our case, an execution cycle of the environment is triggered when a message of our remote agent arrives via the Web socket connection.
\begin{minted}[fontsize=\scriptsize]{js}
    const runnerGen = ws => run => () => {
      ws.on('message', message => {
        const jMessage = JSON.parse(message)
        if (jMessage.agentId && jMessage.actions)
            global.actionRequests[jMessage.agentId] = jMessage.actions
        run()
      })
    }
\end{minted}
On server side, we now have all components for implementing the environment (and running it).
\begin{minted}[fontsize=\scriptsize]{js}
    const environment = new Environment(
      [paranoid, claustrophobe, porter],
      state,
      update,
      state => console.log(state),
      state => state,
      runnerGen(ws)
    )
    environment.run()
\end{minted}
Finally, we implement the client agent.
Again, we abstract from agent-internal details, assuming we have a \verb|claustrophobe| agent, whose reasoning loop we can execute.

To enable the Web socket server integration, we instantiate a client and send an initial opening request to the server to trigger the execution of the environment's loop.
\begin{minted}[fontsize=\scriptsize]{js}
    ws.on('open', () => {
        ws.send(JSON.stringify({ agentId, actions: [[]] }))
    })
\end{minted}
Finally, we implement the Web socket listener that receives messages from the environment and triggers the agent's reasoning loop using these messages as the percepts.
\begin{minted}[fontsize=\scriptsize]{js}
    ws.on('message', message => {
      const jMessage = JSON.parse(message)
      if (jMessage === Object(jMessage)) {
        const actions = claustrophobe.next(jMessage)
        ws.send(JSON.stringify({ agentId, actions }))
      }
    })
\end{minted}
We can then start server and client to run the MAS (leading to a loop in which the porter agent iteratively attempts to satisfy the wishes of the two other agents)\footnote{Cf. Appendix Item~\ref{app:distributed-mas}.}.

\section{Applicability in Different Ecosystems}
\label{sec:applications}
%
Given the ubiquity of the language, JavaScript agents can be deployed across a broad range of environments and devices.
Below, we sketch the relevance of, and some example applications for, JavaScript agents in different ecosystems that can inspire more elaborate applications, as well as in-depth research on related open challenges.
\begin{description}
    \item[\textbf{Agents in Web browsers and other client-side environments.}]
    The most classical deployment target for JavaScript agents is the Web browser. We have show-cased a simple application in Section~\ref{sec:agents}.
    In more serious application scenarios, JavaScript agents and MAS may, for example, run in client environments in order to execute simulations that run directly on user devices, thus off-loading resource consumption from server to client, and potentially facilitating share-ability, e.g., via Web links.
    An example of such an agent-based simulation is presented in~\cite{DBLP:conf/atal/MuallaKTNGN20,DBLP:journals/ai/MuallaTKNCAGN22} (implemented in JS-son); the simulation has been used in a human-computer interaction study.
    Analogously, JavaScript agents can be deployed to desktop and mobile applications using frameworks like Electron\footnote{\url{https://github.com/electron/electron}, \emph{accessed at 2025-02-13}.} and Ionic\footnote{\url{https://github.com/ionic-team/ionic-framework}, \emph{accessed at 2025-02-13.}}, respectively.
    \item[\textbf{Agents in Node.js.}]
    The implementation of the \emph{porter} and \emph{LLM} examples in Sections~\ref{sec:agents}~and~\ref{sec:mas} run on Node.js, which is frequently used for executing JavaScript on Web servers (or as a build tool). More broadly, with Node.js, JavaScript agents can be deployed to most servers and personal computing devices, e.g. devices that run Windows, MacOS, or common Linux distributions.
    \item[\textbf{Agents in Jupyter notebooks.}] With the rise of data science as a practical discipline, technologies that blend programming and data analysis work, such as Jupyter notebooks, have become important tools for scientists as well as for data analysts in industry.
    An example of a simple integration of JS-son-based JavaScript agents and Jupyter notebooks is presented in~\cite{DBLP:conf/emas/KampikN19}, highlighting the interplay of Python and JavaScript\footnote{Cf. Appendix Item~\ref{app:jupyter}.}. Future work could investigate more advanced use cases involving additional, mature technologies for fully integrating JavaScript into data science environments, such as Deno\footnote{\url{https://docs.deno.com/runtime/reference/cli/jupyter/}, \emph{accessed at 2025-02-13.}}.
    \item[\textbf{Agents in Function-as-a-Service environments.}]
    JavaScript is frequently used in Function-as-a-Service (FaaS) environments.
    FaaS offerings facilitate the development and deployment of micro-services by allowing developers to run relatively small and often stateless programs in the cloud with minimal management and maintenance overhead, as the FaaS provider takes care of all abstractions below business logic level (as well as of infrastructure).
    An example tutorial showcasing agents---in this case a simple agent-based simulation---that can de deployed to FaaS environments is available in the JS-son project repository\footnote{Cf. Appendix Item~\ref{app:serverless}.}.
    \item[\textbf{LLM agents.}]
    One of the examples presented in Section~\ref{sec:agents} makes use of an LLM, thus relating to the emerging research trend of LLM agents, i.e., the integration of reasoning loop agents (in the broader sense) and LLMs.
    Arguably, common narratives around LLM agents in academia and industry pay little attention to the history of research on reasoning loop agents and AOP.
    Accordingly, there may be a research gap between LLM agent design patterns and long-running research on AOP. Future research could analyze this gap comprehensively, potentially identifying ways in which AOP can address challenges of LLM agents. In this context, pragmatic, practice-oriented reasoning loop approaches like the ones presented in this chapter may be a first, preliminary starting point for bridging the gap.
    \item[\textbf{Towards agents on constrained devices.}]
    A generally acknowledged research challenge at the intersection of AI and software systems engineering is moving somewhat ``intelligent'' behavior to the edge of networks, i.e., closer to the problems at hand but also to less resourceful, \emph{constrained} computing devices. 
    In this context, research has explored the deployment of agents and use of AOP~\cite{10.1007/978-3-030-66412-1_22,DBLP:journals/corr/abs-2406-17303}.
    Indeed, agents implemented using a slimmed-down version of JS-son have experimentally been deployed to \emph{Espruinos}, microcontroller-based devices that support the execution of JavaScript~\cite{DBLP:conf/atal/KampikGCM21}. 
    However, JS-son is not optimized for performance and resource utilization.
    Indeed, one can intuitively gauge that JS-son's reasoning loop cycle with its nested and otherwise intertwined function calls can lead to inefficient programs. This is less of a problem in environments such as Web browsers, where resources are abundant and efficiency is, commonly, an afterthought.
    In contrast, on constrained devices, JavaScript agents compete with program code in lower-level languages, thus raising the bar in terms of performance requirements.
\end{description}
%
\section{Conclusions}
\label{sec:conclusion}
%
In this chapter, we have provided an overview of how to implement reasoning loop agents and multi-agent systems in JavaScript.
We have highlighted how such JavaScript agents can be implemented with relative ease, either from scratch in vanilla JavaScript, or using \emph{JS-son}, a small agent programming library.
Such agents can then be straightforwardly deployed to a range of technology ecosystems, such as Web front-ends, Node.js back-ends and Juypter notebook-based data science environments.
We hope that our guide provides a useful starting point for implementing agents and MAS in JavaScript, for researchers working on the foundations of agent technologies, applied scientists and practitioners who implement agents and MAS for practical purposes, as well as for educators teaching agent and multi-agent systems to diverse student groups.
Specifically, yet not exhaustively, we recommend considering the application of JavaScript and JS-son agents in the following broad scenarios.
\begin{description}
    \item[\textbf{Teaching AOP to non-computer science students.}] Traditional AOP languages, such as AgentSpeak~\cite{DBLP:conf/maamaw/Rao96}, adopt the logic programming paradigm, which arguably makes them difficult to learn for beginners who lack computer science knowledge. One can reasonably assume that JavaScript (or, alternatively, Python) is more accessible for ``non-technical'' students, who may, for example, start learning about artificial intelligence primarily from the perspectives of the social sciences and humanities, i.e., from disciplines such as philosophy and psychology. While Python may have emerged as the programming language of choice in such contexts, having another mainstream language available for teaching purposes not only allows broadening \emph{polyglot} perspectives on programming languages but also facilitates the implementation of agents and MAS with a focus on different use cases, such as in \emph{Web} contexts where the choice of Python as a programming language is less natural.
    \item[\textbf{Creating share-able agent-based simulations.}] A key advantage of JavaScript-based simulations implemented with libraries like JS-son is that they can be deployed as ``static'' Web pages that do not require any back-end (neither a server, nor a ``serverless'' environment). This facilitates sharing of simulations (and entire simulators), for example as a means to provide interactive science demonstrators using agent technologies~\cite{DBLP:conf/atal/MuallaKTNGN20}.
    \item[\textbf{Implementing light-weight agents with heterogeneous deployment targets.}] As demonstrated above, JavaScript reasoning loop agents can be deployed to a broad range of environment types, such as function-as-a-service providers, Web front-ends, and (possibly constrained) IoT devices. Use cases that require the deployment of conceptually somewhat similar agents to a range of these (and other) environment types may benefit from JavaScript and JS-son agents that can be easily moved between environments, potentially even at run-time.
\end{description}
Beyond these use-case suggestions for JavaScript agents, the conceptual part of this chapter may provide a starting point for closing the gap between reasoning loop agents in the traditional sense and the pragmatic implementation of agents in mainstream programming languages and software technology ecosystems.
Still, the long-term ambition to establish agent-oriented abstractions in the mainstream of programming fundamentals requires continuous efforts.
Accordingly, we consider the following two lines of future work particularly relevant.
\begin{description}
    \item[\textbf{Engineering design patterns for reasoning loop agents.}] With the emergence of LLM agents, \emph{agent design patterns} have emerged as important practical abstractions\footnote{Cf. \url{https://www.anthropic.com/research/building-effective-agents}, \emph{accessed at 2025-02-21}.}. An apparent advantage of these design patterns is their focus on easy-to-understand, generally applicable and purely conceptual/informal abstractions, and the resulting broad appeal to software engineers and even non-programmers. These patterns tend to be detached from the academic AOP literature. Future research can explore \emph{i)} to what extent similarly intuitive, high-level design patterns can be compiled from the AOP literature and \emph{ii)} how the AOP literature can contribute to LLM agent design patterns and, more broadly, ground the current hype around agentic AI.
    In this context, one could set out to refine \emph{design patterns} for reasoning loop agents in a language-agnostic manner, targeted towards practical desiderata.
    \item[\textbf{Software frameworks for industry-scale JavaScript agents.}] The concepts and abstractions for implementing JavaScript agents presented in this chapter are rooted in engineering intuition and pragmatism.
    Still, some design choices may stray too far from what is considered \emph{idiomatic} in either JavaScript and AOP, and can thus benefit from further refinement.  
    On the entirely technology-oriented side, future work may set out to further mature so far somewhat ``academic'' frameworks and libraries.
    For example, improvements to JS-son---or to entirely novel JavaScript-based agent programming libraries---may provide better support for asynchronous behaviors within reasoning loops and could be optimized for modern tools and build systems that utilize JavaScript extensions such as TypeScript.
\end{description}
We hope that our practice-oriented overview of how to implement reasoning loop agents in JavaScript inspires students, researchers, and practitioners to try out agent-oriented abstractions in JavaScript, and vice versa, JavaScript as a language for the agent programming tool-box.
Ultimately, we would wish that others develop their own approaches and idioms, and ideally even advance the research directions sketched above.

\begin{acknowledgement}
\label{sec:acks}
The author thanks the many researchers in the agents and multi-agent systems community who have helped with feedback and discussions about engineering agents and multi-agent systems, in JavaScript, on the Web, and beyond.
This work was partially supported by the Wallenberg AI, Autonomous Systems and Software Program (WASP) funded by the Knut and Alice Wallenberg Foundation.
\end{acknowledgement}

\section*{Appendix}
The appendix contains links to code examples and documentation, sometimes accompanied by brief remarks.
\begin{enumerate}
    \item \textbf{JS-son Documentation}\label{app:js-son-doc} \\
    The JS-son documentation is available at \url{https://js-son.readthedocs.io/en/latest/}, \emph{accessed at 2025-02-12}.
    \item \textbf{Jason Room Example}\label{app:jason-room} \\
    Our example is based on one of the standard examples of the seminal \emph{Jason} AgentSpeak interpreter: 
    \url{https://github.com/jason-lang/jason/tree/main/examples/room}, \emph{accessed at 2025-12-02}. Our simple example here is a partial, single-agent implementation.
    \item \textbf{Vanilla JavaScript Agent}\label{app:vanilla-example} \\
    The example is, with additional console log statements for the sake of illustration, available at \url{https://gist.github.com/TimKam/2825ff18367090eb2e9bbd4db224723c} (\emph{accessed at 2025-02-18}) and can, for instance, be executed by pasting the entire content of the corresponding file into a Web browser console.
    \item \textbf{JS-son LLM Agent}\label{app:llm-agent} \\
    The example's code is available at \url{https://github.com/TimKam/JS-son/tree/master/examples/llm}, \emph{accessed at 2025-02-19}.
    \item \textbf{JS-son Centralized MAS}\label{app:centralized-mas} \\
    The example's code, also covering the browser-based implementation, is available at \url{https://github.com/TimKam/JS-son/tree/master/examples/web}, \emph{accessed at 2025-02-14}.
    \item \textbf{JS-son Distributed MAS}\label{app:distributed-mas} \\
    The example's code (including all agent and environment internals that we skip in the walk-through we provide in this chapter) is available at \url{https://github.com/TimKam/JS-son/tree/master/examples/distributed}, \emph{accessed at 2025-02-18}.
    \item \textbf{JS-son Agents and Jupyter Notebooks}\label{app:jupyter}
    The example's code is available at \url{https://github.com/TimKam/JS-son/tree/master/examples/jupyter}, \emph{accessed at 2025-02-13}.
    \item \textbf{Severless JS-son Agents}\label{app:serverless}
    The example's code is available at \url{https://github.com/TimKam/JS-son/tree/master/examples/serverless}, \\ \emph{accessed at 2025-02-12}.
\end{enumerate}

%
%
%
\bibliographystyle{spmpsci}
\bibliography{references}

\begin{thebibliography}{10}
\providecommand{\url}[1]{{#1}}
\providecommand{\urlprefix}{URL }
\expandafter\ifx\csname urlstyle\endcsname\relax
  \providecommand{\doi}[1]{DOI~\discretionary{}{}{}#1}\else
  \providecommand{\doi}{DOI~\discretionary{}{}{}\begingroup \urlstyle{rm}\Url}\fi

\bibitem{DBLP:conf/eumas/BaiardiBCP23}
Baiardi, M., Burattini, S., Ciatto, G., Pianini, D.: Jakta: {BDI} agent-oriented programming in pure kotlin.
\newblock In: V.~Malvone, A.~Murano (eds.) Multi-Agent Systems - 20th European Conference, {EUMAS} 2023, Naples, Italy, September 14-15, 2023, Proceedings, \emph{Lecture Notes in Computer Science}, vol. 14282, pp. 49--65. Springer (2023).
\newblock \doi{10.1007/978-3-031-43264-4\_4}.
\newblock \urlprefix\url{https://doi.org/10.1007/978-3-031-43264-4\_4}

\bibitem{BOISSIER2013747}
Boissier, O., Bordini, R.H., HÃŒbner, J.F., Ricci, A., Santi, A.: Multi-agent oriented programming with jacamo.
\newblock Science of Computer Programming \textbf{78}(6), 747 -- 761 (2013).
\newblock \doi{https://doi.org/10.1016/j.scico.2011.10.004}.
\newblock \urlprefix\url{http://www.sciencedirect.com/science/article/pii/S016764231100181X}.
\newblock Special section: The Programming Languages track at the 26th ACM Symposium on Applied Computing (SAC 2011) \& Special section on Agent-oriented Design Methods and Programming Techniques for Distributed Computing in Dynamic and Complex Environments

\bibitem{Bordini:2007:PMS:1197104}
Bordini, R.H., H\"{u}bner, J.F., Wooldridge, M.: Programming Multi-Agent Systems in AgentSpeak Using Jason (Wiley Series in Agent Technology).
\newblock John Wiley \& Sons, Inc., USA (2007)

\bibitem{rfc8259}
Bray, T.: {The JavaScript Object Notation (JSON) Data Interchange Format}.
\newblock RFC 8259 (2017).
\newblock \doi{10.17487/RFC8259}.
\newblock \urlprefix\url{https://www.rfc-editor.org/info/rfc8259}

\bibitem{DBLP:conf/emas/BriolaFM23}
Briola, D., Ferrando, A., Mascardi, V.: Fantastic mass and where to find them: First results and lesson learned.
\newblock In: A.~Ciortea, M.~Dastani, J.~Luo (eds.) Engineering Multi-Agent Systems - 11th International Workshop, 2023, London, UK, May 29-30, 2023, Revised Selected Papers, \emph{Lecture Notes in Computer Science}, vol. 14378, pp. 233--252. Springer (2023).
\newblock \doi{10.1007/978-3-031-48539-8\_16}.
\newblock \urlprefix\url{https://doi.org/10.1007/978-3-031-48539-8\_16}

\bibitem{programminglanguages}
Cass, S.: The top programming languages 2024.
\newblock IEEE Spectrum  (2024).
\newblock \urlprefix\url{https://spectrum.ieee.org/ibm-quantum-computer-2668978269}

\bibitem{DBLP:conf/emas/CiorteaBR18}
Ciortea, A., Boissier, O., Ricci, A.: Engineering world-wide multi-agent systems with hypermedia.
\newblock In: D.~Weyns, V.~Mascardi, A.~Ricci (eds.) Engineering Multi-Agent Systems - 6th International Workshop, {EMAS} 2018, Stockholm, Sweden, July 14-15, 2018, Revised Selected Papers, \emph{Lecture Notes in Computer Science}, vol. 11375, pp. 285--301. Springer (2018).
\newblock \doi{10.1007/978-3-030-25693-7\_15}

\bibitem{gradner1970fantastic}
Gradner, M.: The fantastic combinations of john conway's new solitaire game life.
\newblock Scientific American \textbf{223}(4), 120--123 (1970)

\bibitem{DBLP:conf/atal/KampikGCM21}
Kampik, T., Gomez, A., Ciortea, A., Mayer, S.: Autonomous agents on the edge of things.
\newblock In: F.~Dignum, A.~Lomuscio, U.~Endriss, A.~Now{\'{e}} (eds.) {AAMAS} '21: 20th International Conference on Autonomous Agents and Multiagent Systems, Virtual Event, United Kingdom, May 3-7, 2021, pp. 1767--1769. {ACM} (2021).
\newblock \doi{10.5555/3463952.3464231}.
\newblock \urlprefix\url{https://www.ifaamas.org/Proceedings/aamas2021/pdfs/p1767.pdf}

\bibitem{DBLP:conf/emas/KampikN19}
Kampik, T., Nieves, J.C.: Js-son - {A} lean, extensible javascript agent programming library.
\newblock In: L.A. Dennis, R.H. Bordini, Y.~Lesp{\'{e}}rance (eds.) Engineering Multi-Agent Systems - 7th International Workshop, {EMAS} 2019, Montreal, QC, Canada, May 13-14, 2019, Revised Selected Papers, \emph{Lecture Notes in Computer Science}, vol. 12058, pp. 215--234. Springer (2019).
\newblock \doi{10.1007/978-3-030-51417-4\_11}

\bibitem{engineering-gsi-article-2019}
Mascardi, V., Weyns, D., Ricci, A., Earle, C.B., Casals, A., Challenger, M., Chopra, A., Ciortea, A., Dennis, L.A., D{\'i}az, {\'A}.F., Fallah-Seghrouchni, A.E., Ferrando, A., Fredlund, L.{\AA}., Giunchiglia, E., Guessoum, Z., G{\"u}nay, A., Hindriks, K., Iglesias, C.A., Logan, B., Kampik, T., Kardas, G., Koeman, V.J., Larsen, J.B., Mayer, S., M{\'e}ndez, T., M{\'e}ndez, T., Nieves, J.C., Seidita, V., Tezel, B.T., Varga, L.Z., Winikoff, M.: {E}ngineering {M}ulti-{A}gent {S}ystems: {S}tate of {A}ffairs and the {R}oad {A}head.
\newblock SIGSOFT Engineering Notes (SEN)  (2019)

\bibitem{DBLP:conf/atal/MuallaKTNGN20}
Mualla, Y., Kampik, T., Tchappi, I.H., Najjar, A., Galland, S., Nicolle, C.: Explainable agents as static web pages: {UAV} simulation example.
\newblock In: D.~Calvaresi, A.~Najjar, M.~Winikoff, K.~Fr{\"{a}}mling (eds.) Explainable, Transparent Autonomous Agents and Multi-Agent Systems - Second International Workshop, {EXTRAAMAS} 2020, Auckland, New Zealand, May 9-13, 2020, Revised Selected Papers, \emph{Lecture Notes in Computer Science}, vol. 12175, pp. 149--154. Springer (2020).
\newblock \doi{10.1007/978-3-030-51924-7\_9}.
\newblock \urlprefix\url{https://doi.org/10.1007/978-3-030-51924-7\_9}

\bibitem{DBLP:journals/ai/MuallaTKNCAGN22}
Mualla, Y., Tchappi, I., Kampik, T., Najjar, A., Calvaresi, D., Abbas{-}Turki, A., Galland, S., Nicolle, C.: The quest of parsimonious {XAI:} {A} human-agent architecture for explanation formulation.
\newblock Artif. Intell. \textbf{302}, 103573 (2022).
\newblock \doi{10.1016/J.ARTINT.2021.103573}.
\newblock \urlprefix\url{https://doi.org/10.1016/j.artint.2021.103573}

\bibitem{DBLP:conf/emas/ONeillC23}
O'Neill, E., Collier, R.W.: Exploiting service-discovery and openapi in multi-agent microservices {(MAMS)} applications.
\newblock In: A.~Ciortea, M.~Dastani, J.~Luo (eds.) Engineering Multi-Agent Systems - 11th International Workshop, {EMAS} 2023, London, UK, May 29-30, 2023, Revised Selected Papers, \emph{Lecture Notes in Computer Science}, vol. 14378, pp. 78--84. Springer (2023).
\newblock \doi{10.1007/978-3-031-48539-8\_5}.
\newblock \urlprefix\url{https://doi.org/10.1007/978-3-031-48539-8\_5}

\bibitem{DBLP:journals/corr/abs-2406-17303}
Ramanathan, G., Gomez, A., Mayer, S.: Learnings from implementation of a {BDI} agent-based battery-less wireless sensor.
\newblock CoRR \textbf{abs/2406.17303} (2024).
\newblock \doi{10.48550/ARXIV.2406.17303}.
\newblock \urlprefix\url{https://doi.org/10.48550/arXiv.2406.17303}

\bibitem{DBLP:conf/maamaw/Rao96}
Rao, A.S.: Agentspeak(l): {BDI} agents speak out in a logical computable language.
\newblock In: W.V. de~Velde, J.W. Perram (eds.) Agents Breaking Away, 7th European Workshop on Modelling Autonomous Agents in a Multi-Agent World, Eindhoven, The Netherlands, January 22-25, 1996, Proceedings, \emph{Lecture Notes in Computer Science}, vol. 1038, pp. 42--55. Springer (1996).
\newblock \doi{10.1007/BFB0031845}.
\newblock \urlprefix\url{https://doi.org/10.1007/BFb0031845}

\bibitem{rao91a}
Rao, A.S., Georgeff, M.P.: Modeling rational agents within a {BDI}-architecture.
\newblock In: J.~Allen, R.~Fikes, E.~Sandewall (eds.) Proceedings of the 2nd International Conference on Principles of Knowledge Representation and Reasoning, pp. 473--484. Morgan Kaufmann publishers Inc.: San Mateo, CA, USA (1991)

\bibitem{russel2010}
Russell, S., Norvig, P.: Artificial Intelligence: A Modern Approach, 3 edn.
\newblock Prentice Hall (2010)

\bibitem{SHOHAM199351}
Shoham, Y.: Agent-oriented programming.
\newblock Artificial Intelligence \textbf{60}(1), 51--92 (1993).
\newblock \doi{https://doi.org/10.1016/0004-3702(93)90034-9}.
\newblock \urlprefix\url{https://www.sciencedirect.com/science/article/pii/0004370293900349}

\bibitem{team2023gemini}
Team, G., Anil, R., Borgeaud, S., Alayrac, J.B., Yu, J., Soricut, R., Schalkwyk, J., Dai, A.M., Hauth, A., Millican, K., et~al.: Gemini: a family of highly capable multimodal models.
\newblock arXiv preprint arXiv:2312.11805  (2023)

\bibitem{10.1007/978-3-030-66412-1_22}
Vente, S., Kimmig, A., Preece, A., Cerutti, F.: Increasing negotiation performance at the edge of the network.
\newblock In: N.~Bassiliades, G.~Chalkiadakis, D.~de~Jonge (eds.) Multi-Agent Systems and Agreement Technologies, pp. 351--365. Springer International Publishing, Cham (2020)

\bibitem{llmagentsurvey}
Wang, L., Ma, C., Feng, X., Zhang, Z., Yang, H., Zhang, J., Chen, Z., Tang, J., Chen, X., Lin, Y., Zhao, W.X., Wei, Z., Wen, J.: A survey on large language model based autonomous agents.
\newblock Frontiers of Computer Science \textbf{18}(6), 186345 (2024).
\newblock \doi{10.1007/s11704-024-40231-1}.
\newblock \urlprefix\url{https://doi.org/10.1007/s11704-024-40231-1}

\end{thebibliography}
\end{document}